\documentstyle[aasms4]{article}
\begin{document}
\title{Sgr A$^*$: A supermassive black hole or a spatially extended 
object?}
\author{F. Munyaneza, D. Tsiklauri\footnote{email: 
tsiklauri@physci.uct.ac.za, http://pc021.phy.uct.ac.za/tsiklauri/}, 
and R.D. Viollier\footnote{email: viollier@physci.uct.ac.za}}
\affil{Physics Department, University of Cape Town, Rondebosch 7701,
South Africa}
\begin{abstract}
We report here on a  calculation of possible
orbits of the fast moving infrared source S1 
which has been recently observed  by Eckart \& Genzel (1997)
near the Galactic center. It is shown that tracking of the orbit 
of S1 or any other fast moving star near Sgr A$^*$ offers
a possibility of distinguishing between the supermassive
black hole and extended object scenarios of Sgr A$^*$.
In our calculations we assumed that the extended object
at the Galactic center is a non-baryonic ball made of
degenerate, self-gravitating heavy neutrino matter, as
it has been recently proposed by Tsiklauri \& Viollier (1998a,b).  
\end{abstract}

In spite of the vast and tantalizing work undertaken to 
resolve the issue of the
existence of supermassive black holes (BH) at the centers of 
galaxies, it seems fair to say that, in this case,
the jury is still out (for a current status see e.g. a 
review paper by Ho 1998). 
The discovery of quasars in the early 1960's provoked the
idea that these extremely powerful emission sources 
draw their energy from  accretion of matter onto
compact, supermassive objects of $10^6$ to   $10^9$ solar
masses. This has led many to believe that these objects
are supermassive black holes
(Zeldovich \& Novikov 1964; Salpeter 1964; Lynden-Bell 1969). 
However, as it has also often been pointed out 
(e.g. Kormendy \& Richstone 1995, Maoz 1998), 
the current belief
of the scientific community that the driving engines of 
AGNs are actually supermassive black holes 
largely rests on the {\it implausibility} of
alternative explanations, in particular, explanations 
which are based  on some form of clustered {\it baryonic}
matter. In this context, it is  worthwhile to recall
so-called "totalitarian principle of physics"
(Ostriker, 1971). This principle, which was originally formulated
by the late British novelist T.H. White as the governing
principle for a colony of ants, states that "anything
not forbidden is compulsory."
It has been used to advocate many speculative
(at those days) ideas both in particle physics
and in astrophysics. For example,
the possible existence of neutron stars had been predicted
in the early 1930's, however until they were actually
detected in the late 1960's, they seemed to be too
exotic to be credible
for the scientific community at large. 
Thus, in order to test the
supermassive BH paradigm, acceptable alternative
models for objects such as Sgr A$^*$ ought to be developed.  
In fact, we have shown
(Tsiklauri \& Viollier 1998a) that the matter 
concentration observed through stellar motion at the Galactic center
(Eckart \& Genzel 1997 and
Genzel et al. 1996) is consistent with a 
supermassive object of $2.5 \times 10^6$  solar masses 
consisting of
self-gravitating, degenerate heavy neutrino matter.
While in the limit of neutrino masses of a few hundreds
of keV$/c^2$, which corresponds to the Oppenheimer-Volkoff
limit (Bili\'c at al. 1998), the BH and neutrino ball
scenarios are virtually indistinguishable, they differ
substantially for neutrino masses between 10 to 20 keV$/c^2$.
In fact, the observational data (Eckart \& Genzel 1997 and
Genzel et al. 1996) on the compact dark object
at the Galactic center restrict the neutrino mass to 
$m_\nu \geq 12.0$ keV$/c^2$
for $g=2$ or $m_\nu \geq 14.3$ keV$/c^2$ for $g=1$,
where $g$ is the spin degeneracy factor of the neutrino.
Using these lower bounds, an acceptable fit to the
infrared and part of the radio spectrum emitted by Sgr A$^*$ was 
obtained in the framework of the standard accretion disk theory 
(Tsiklauri \& Viollier 1998b).

The  most reliable techniques of deducing the masses 
of central objects
in galaxies are based on  stellar proper motions, since 
stars are direct tracers of the gravitational potential and
are not affected by non-gravitational forces, in contrast to 
gas clouds, for example, which 
may be vastly affected by existing magnetic fields. 
Even the most 
sophisticated techniques developed so far,
where the mass of central objects is determined
using the first moment of the collisionless 
Boltzmann equation 
(also referred to as Jeans equation, Binney \& Tremaine 1987;
Tsiklauri \& Viollier 1998a), are model {\it dependent}. 
Moreover, as was pointed out by Kormendy \& Richstone 1995,
the results obtained in this approach are quite sensitive to the
effects of velocity anisotropy.  
In fact, an increase
of the velocity dispersion towards the center as
$\propto{r}^{-1/2}$, 
which, assuming an isotropic velocity distribution could serve
as proof of the point-likeness of the central object,
can effectively be mimicked by an anisotropic velocity distribution.
Although, some progress has been made in removing this degeneracy
(see Ho, 1998 for details), there is still ample 
room for speculations,
which  models like our own (Tsiklauri \& Viollier 1998a,b) 
try to exploit.
While at projected distances from Sgr A$^*$ larger than 
0.1 pc the number of stars observed within a shell of a 
given projected thickness is large enough to make
the calculation of the velocity dispersion statistically
sound, this number drops to only 11 stars (S1-S11)
within 0.03 pc, rendering a statistical treatment of the 
innermost stars around Sgr A$^*$ somewhat meaningless.
Thus, in this letter we would like to explore the gravitational
potential of Sgr A$^*$ without using statistics, i.e. studying
the motion of individual stars in the vicinity of Sgr A$^*$.
Such an analysis has the   
advantage of being model {\it independent}, as 
our arguments will be solely based on Newtonian dynamics.
The idea of tracking individual stars
on an appreciable fraction of the orbiting period,
in order to distinguish between the supermassive 
black hole and extended object or neutrino ball scenarios of
Sgr A$^*$, was proposed in Tsiklauri \& Viollier 1998a. 

As an example we investigate here the motion of the
fastest (S1) of the eleven stars in the central cluster around
Sgr A$^*$. 
We study the dynamics of the star S1 for two distinct cases:
first, assuming that Sgr A$^*$ is a black hole of the mass 
$2.61 \times 10^6 M_\odot$ and second, replacing the 
supermassive black hole by an extended object, i.e.,
a ball of the same mass, consisting of degenerate, 
self-gravitating neutrino (and antineutrino) matter.
In a previous paper, Tsiklauri \& Viollier 1998a
have set bounds on the mass of the neutrino, based on the
observational data taken form Genzel et al. 1996.
Since then, the data points describing the mass enclosed
within a given radius have moved inwards,
thus putting new constraints on the possible size of the central
object.  
It is therefore worthwhile to reconcile our model with the most 
recent published data by Genzel et al. 1997,
who  established that the enclosed mass at the Galactic
center within 0.016 pc (the innermost data point)
is $(2.65 \pm 0.76)\times 10^6 M_\odot$.
Repeating the analysis of Tsiklauri \& Viollier 1998a, 
using these new data, we find that the
new bounds on the mass of the neutrino are $m_\nu \geq 12.07$ keV$/c^2$
for $g=2$ or $m_\nu \geq 14.35$ keV$/c^2$ for $g=1$,
which slightly reduce the maximal radius of the neutrino ball.
Thus, using the value of $2.61 \times 10^6 M_\odot$
(Genzel et al. 1997) for the mass of the neutrino ball,
the radius turns out to be $2.477 \times 10^{-2}$ pc (we assume 
that the distance to Sgr A$^*$ is 8.0 kpc throughout this letter).

Eckart \& Genzel 1997 have 
presented proper motion data for S1, as well
as other stars in the central cluster around Sgr A$^*$.
In 1994 the coordinates of S1 were 
RA=-0.19$''$  and DEC=-0.04$''$, with  Sgr A$^*$ being
the origin of the coordinate system, and the
x- and y-components of the projected velocity,
$v_x=650 \pm 400$ km/sec and $v_y=-1530 \pm 400$ km/sec, 
respectively,
deduced from the 1994 and 1996 data. Here $x$ is opposite to the
RA direction and $y$ is in DEC direction.
We then solve Newton's equations for two cases (i)
a black hole with a mass
$2.61 \times 10^6 M_\odot$ and (ii) a neutrino ball of the same
mass, with the neutrino mass fixed at the lower limit allowed
by the stellar proper motion data (Genzel et al. 
1997). Note, that increasing of the neutrino mass will
smoothly interpolate between the scenarios (ii) and (i). 
A typical result of such calculation is shown in
Fig. 1 where we plot the two orbits of S1 corresponding to the BH
and neutrino ball scenarios. The neutrino ball is represented by
the dotted circle with its center (star) at the position 
of Sgr A$^*$.
The values for $v_x$ and $v_y$ are taken as 650 and $-1530$ km/sec,
respectively. 
Sgr A$^*$ and S1 are assumed to be at the same distance from the
observer, i.e. the $z$ coordinate of the star S1, as measured
in the line-of-sight from Sgr A$^*$, is zero. Moreover, the velocity
component  in the line-of-sight of the star S1, $v_z$, has also
been set to zero in this figure.
The unknown quantities, $z$ and $v_z$, are
the major source of uncertainty in determining the intrinsic
orbit of the star S1.
However, as we will see below, 
this shortcoming will not substantially affect the predictive
power of our model if appropriately dealt with.
Finally, the full square labels on the orbits denote time
in years. Due to the fact that the gravitational force
at a given distance from Sgr A$^*$ is determined by the mass
enclosed within this distance, the star S1 will be deflected much
less in the neutrino ball scenario than in the BH scenario
of Sgr A$^*$, as can be seen  from Fig. 1.

It is worthwhile to note at this stage
that the observational test which
we propose is somewhat reminiscent of Rutherford's
experiments at the beginning of this century. 
These experiments led to abandoning of Thomson's "pudding"
model of the atom (which described atom as an extended
positively charged spherical cloud, with electrons
like raisins in a pudding whose oscillation around the
equilibrium point was providing electromagnetic radiation)
and established the current views of the atomic structure of
matter and the "compactness" of the nucleus.

Returning back to  Fig.1, it seems that,
since the positions of the stars S1-S11 are known
to 30 mas accuracy, distinguishing the BH from 
the neutrino ball
scenario for Sgr A$^*$ might be possible in a few years time.
However, as we will discuss  below, this estimate is perhaps 
too optimistic,
since it does not take into account the uncertainties 
due to the complete lack of information on $z$ and $v_z$.
Moreover, as we know, there are also large uncertainties in 
the determination of $v_x$ and $v_y$. 
Of course, all these uncertainties will eventually decrease,
as more data will become available, since the projected
orbit, inclusive $v_z$ and $z$, will be completely
determined  by the accurate measurement of  the
position of S1 as a function of time.
Thus, as our next step,
we will investigate the errors in the velocity
components in more detail. To this end,
we have performed  calculations of the 
orbit of the star S1  in both
BH and neutrino ball scenarios, taking into 
account the error bars
of $v_x$ and $v_y$. The results of this calculation are
presented in the Fig. 2. The top panel represents 
the orbits  in the case of a BH, whereas, the lower
panel describes the  neutrino ball scenario.
The spread of the orbits  induced by the uncertainties in
$v_x$ and $v_y$ is quite large. 
In the BH scenario, the
curves 1, 2 and 4 are bound orbits (ellipses with the 
BH in one focus),
whereas in the neutrino ball scenario only one orbit is
bound (curve 2, which is actually a rosette-type orbit). 
In both cases we assume  $z=v_z=0$.
At first sight, it may seem
that the results of this calculation are inconclusive,
especially in view of the current complete uncertainty in
$z$ and $v_z$.
However, let us investigate in some more detail the dependence
of the orbits presented in the Fig. 2 on these two 
parameters. In Fig. 3 where we plot the orbits of the 
star S1
for $v_z=0$, $v_x$ and $v_y$ being 
fixed at 650 and $-1530$ km/sec, respectively, and
varying $z$. The top panel corresponds to the 
BH scenario, whereas
lower one is for the neutrino ball case. 
The largest value of $z$ taken 
in this plot corresponds to the radius 
of the neutrino ball, i.e. the distance
from Sgr A$^*$ beyond which there is, obviously,
no difference between the BH and neutrino ball 
scenarios.
It is worthwhile to note that in the neutrino ball
scenario the dependence on
$z$ is relatively small as soon as $z$ is smaller
than its radius.
This is can be understood by the fact that the
potential inside the neutrino ball is approximately
of harmonic oscillator-type, where Newton's equations
decouple in Cartesian coordinates.
We also conclude from the plot that, by increasing 
$z$, the orbits become more like  straight lines, 
i.e. they are shifting towards the lower 
right corner of the graph. 
This is in accordance  to one's expectations,
since increasing of the $z$ coordinate means 
moving away from the scattering center, 
with less interaction and deflection. 

The dependence of the orbits on  $v_z$ is
summarized in Fig. 4 which is similar to Fig. 3, but 
in this case we fix $z$ to zero, 
varying $v_z$ instead. Again, 
increasing $v_z$ produces a  similar effect as
in the case of increasing $z$ in the previous figure, 
i.e. larger $z$-components of the velocity
lead to straightening of the orbit. 
As increasing $v_z$ yields a larger velocity for 
the star,
it is clear that faster moving stars will be 
deflected less than the slower ones.
We have thus established that both uncertainty factors,
$z$ and $v_z$, have similar effects on the orbits. 
In Fig. 5 two orbits are plotted: the upper-leftmost
orbit of S1 corresponding to the neutrino ball scenario
(actually, line 4 in the bottom panel of Fig. 2)
and the nearest orbit of the same object but for the
BH scenario (actually, line 5 in the upper panel of the Fig. 2).
From this figure we can conclude that 
{\it irrespective} of what
the actual values of $z$ and $v_z$ are, 
if the star S1
will eventually be found
in the upper-left zone of the graph, 
then it will clearly rule out the neutrino ball
scenario of Sgr A$^*$ for the chosen neutrino mass. 
This graph
as well as Fig. 2 is plotted for $z=v_z=0$, 
and as we have
learned from Figs. 3 and 4, any non-zero values
of these parameters would just straighten the orbit.
Thus, this plot is a good test for the identification
of a supermassive black hole, as opposed to an 
extended object or a neutrino ball
at the Galactic center. If the orbit of S1
eventually ends up in the lower right corner, then 
Sgr A$^*$ can  be interpreted  as either a neutrino ball
or a BH with a large $z$-parameter. 
Thus, it seems that as the observations proceed
within the next decade, one may tell the difference
between the supermassive 
black hole scenario and neutrino ball model 
of the Galactic center using this test, 
since the uncertainties in the projected orbit will
decrease in the course of time.

Note: After this work has been completed, we became aware of
new results on the central cluster by 
Ghez et al. 1998. While the details of the plots
might change using these new data, the basic results
of this letter will be unaffected.

\newpage
\centerline{figure captions:}
Fig. 1: Projected orbits of the star S1 in the case of a supermassive
BH of a mass $2.61 \times 10^6 M_\odot$ (solid line)
and in the case of a 
neutrino ball (dashed line). The surface of the neutrino ball
with the same  mass  as the BH, and
consisting of heavy, degenerate neutrino matter with neutrino masses  
$m_\nu = 12.07$ keV$/c^2$
for $g=2$ or $m_\nu = 14.35$ keV$/c^2$ for $g=1$
is shown by the dotted circle.
The neutrino ball has a 
radius of $9.9182 \times  10^4$ Schwarzschild radii. 
The star in the upper-right corner
denotes Sgr A$^*$. In this plot 
$v_x=650$ km/sec, $v_y=-1530$ km/sec and $v_z=z=0$.

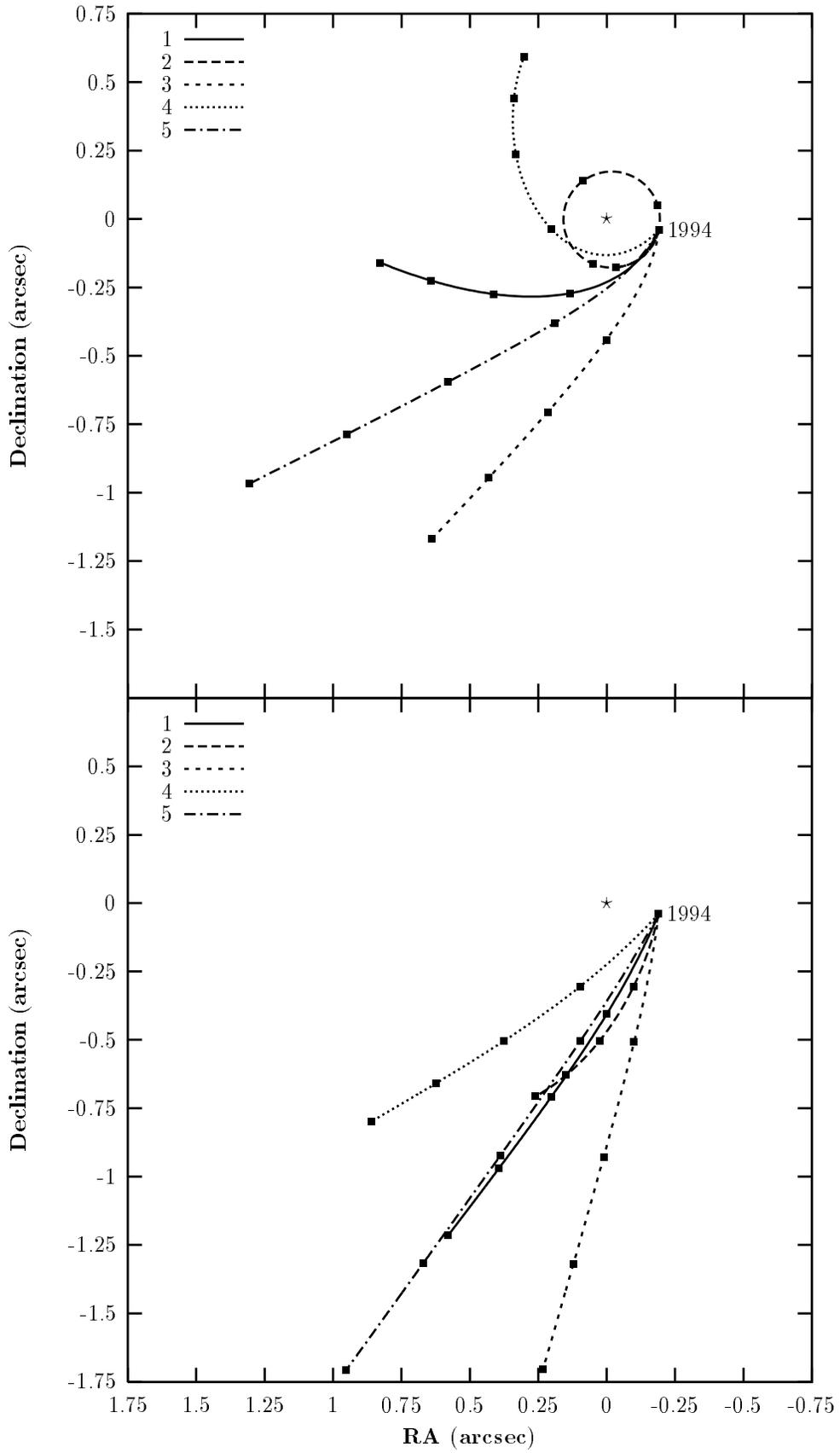
Fig.2: Projected orbits of the star S1 in the case of a supermassive
black hole (top panel)
and in the case of a neutrino ball (lower panel)
for $v_z=z=0$.
In this graph we explore how the orbits  are affected by the
uncertainty in $v_x$ and $v_y$. The labels for orbits are:
1: $v_x=650$ km/sec and $v_y=-1530$ km/sec (median 
values, orbit is bound in the case of a BH and unbound for 
the neutrino ball); 
2: $v_x=250$ km/sec and $v_y=-1130$ km/sec
(orbits are bound in both cases);
3: $v_x=250$ km/sec and $v_y=-1930$ km/sec
(orbits are  unbound in both cases);
4: $v_x=1050$ km/sec and $v_y=-1130$ km/sec
(orbit is bound in the case of a BH and unbound for a neutrino ball);
5: $v_x=1050$ km/sec and $v_y=-1930$ km/sec
(orbits  are unbound in both cases). The 
time labels (filled squares) on the orbits are placed 
in intervals of 10 years, up to the year 2034.

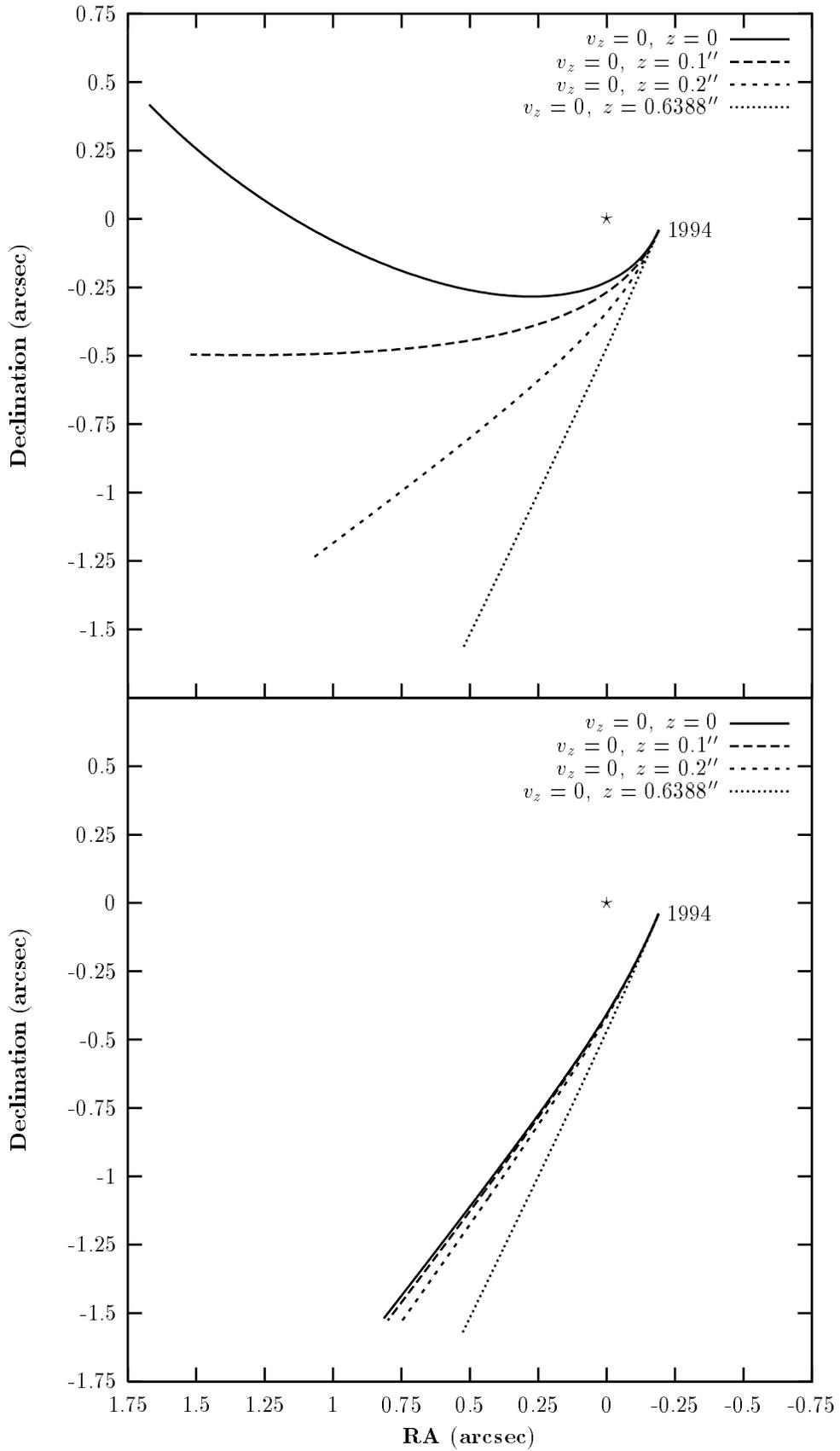
Fig.3:
Projected orbits of the star S1 in the case 
of a supermassive
BH (top panel)
and in the case of a neutrino ball (lower panel).
In this graph we explore how the orbits  are affected by the
uncertainty in the $z$-parameter. The labels for the orbits are 
given in the graph. Note, that for $z=0.6388''$, which
corresponds to the radius of the neutrino ball for the assumed
distance to the Galactic center, the orbits for a BH and
neutrino ball are identical, as it should be.
In this graph $v_x=650$ km/sec, $v_y=-1530$ km/sec and $v_z=0$.

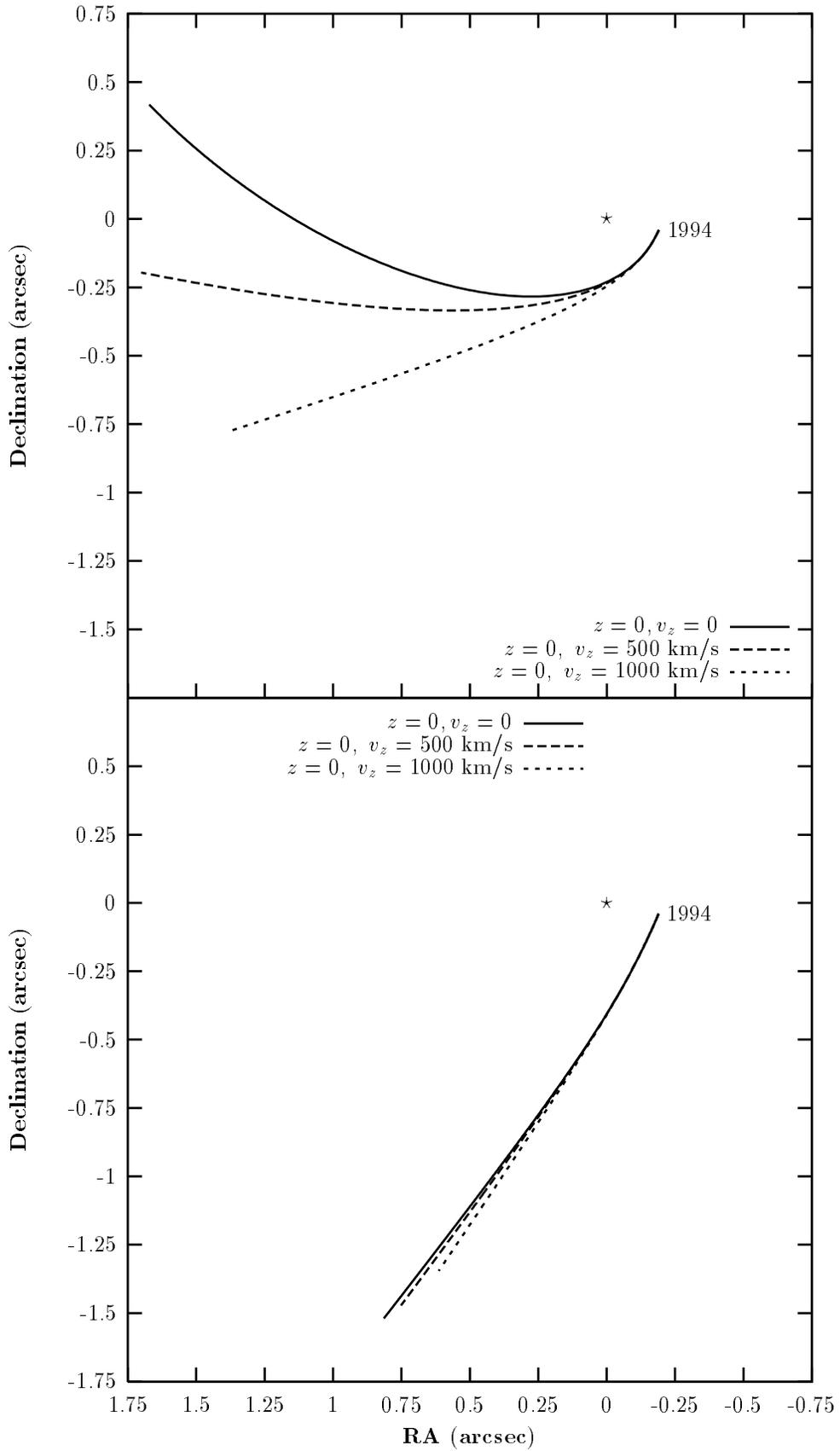
Fig.4:
Projected orbits of the star S1 in the case of a supermassive
black hole (top panel)
and in the case of a neutrino ball (lower panel).
In this graph we explore how the orbits  are affected by the
uncertainty in $v_z$. The labels for the orbits are 
given in the graph. Here, 
 $v_x=650$ km/sec, $v_y=-1530$ km/sec and $z=0$.

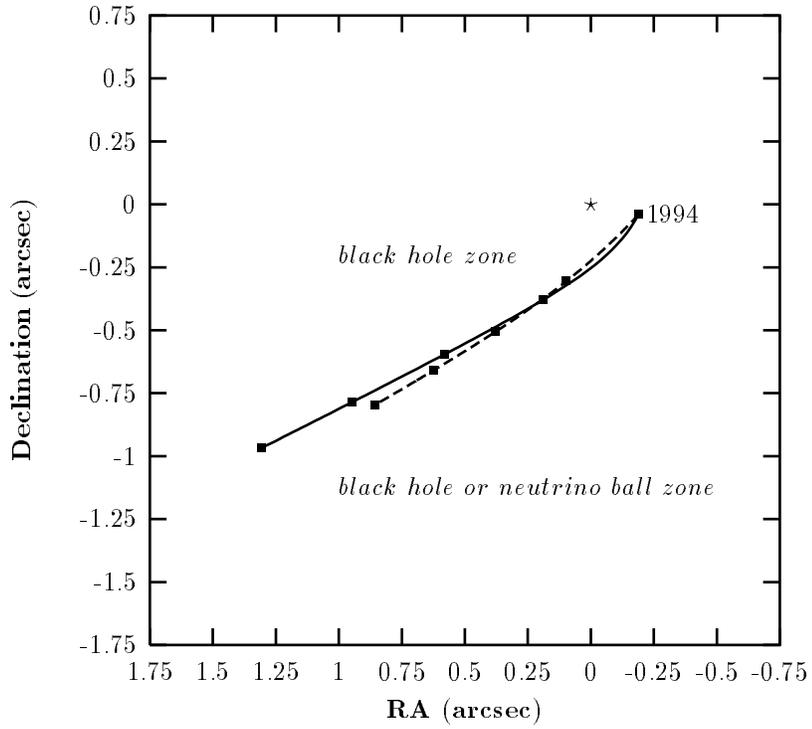
Fig.5:
A graph which combines line 4 from the bottom panel of Fig. 2
and line 5 from the upper panel of Fig. 2 for
comparison. If the star S1 will eventually 
be found in the upper-left zone of the graph, 
i.e. up and left of the
overlapping lines (referred to as a black hole zone), this 
will rule out the neutrino ball 
interpretation of Sgr A$^*$ for the chosen neutrino mass.
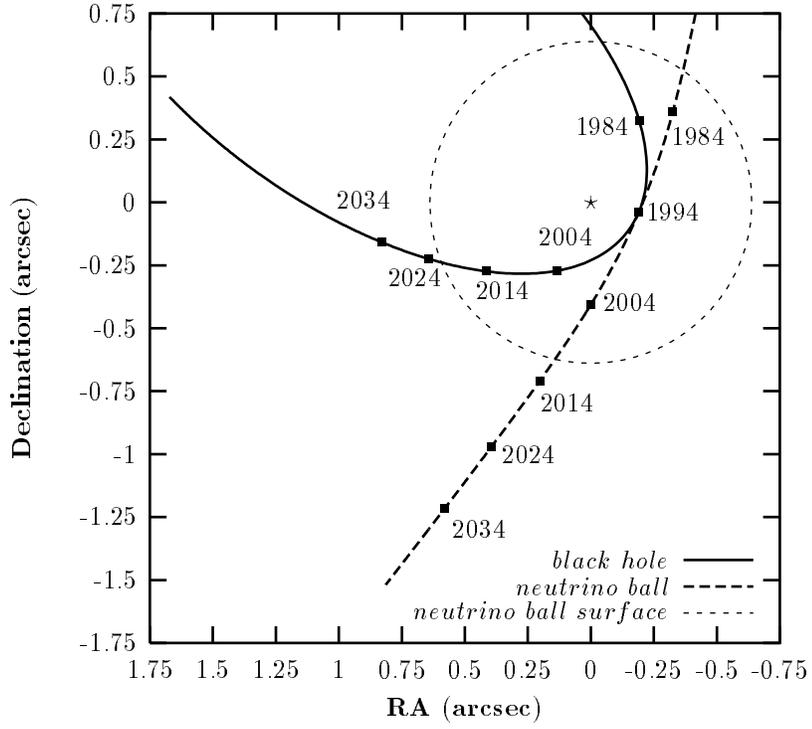
\end{document}